\documentclass[12pt]{article}
\usepackage[dvips]{graphicx}
\usepackage{graphicx}
\usepackage{epsfig}
\usepackage{color}
\usepackage{epstopdf}
\usepackage{amsmath}

\usepackage{caption}
\usepackage{subcaption}
\usepackage{amssymb}
\usepackage{lineno}

\def\be{\begin{equation}}
\def\bea{\begin{eqnarray}}
\def\ee{\end{equation}}
\def\eea{\end{eqnarray}}

\def\eps{\varepsilon}

\def\la{\lambda}

\newcommand{\jgr}{    {J. Geophys. Res.}}
\newcommand{\ssr}{    {Space Sci. Rev.}}

\newcommand{\nat}{ {Nature }}

\def\P{{\cal P}}
\def\la{{\lambda}}
\def\eps{{\varepsilon}}

\begin{document}

\begin{center}
{\Large \bf A map for systems with resonant trappings and scatterings}\\

\vskip 10mm

{\Large  A.~V.~Artemyev$^{1,2}$, A. I. Neishtadt$^{3,2}$, A. A. Vasiliev$^{2}$\\

\vskip 5mm

{\large \it $^{1}$ Institute of Geophysics and Planetary Physics, University of California, Los Angeles, California, USA}\\
{\large \it $^{2}$ Space Research Institute, Moscow, Russia}\\
{\large \it $^{3}$ Department of Mathematical Sciences, \\ Loughborough University, UK}\\
}
\end{center}

\section*{Abstract}
Slow-fast dynamics and resonant phenomena can be found in a wide range of physical systems, including problems of celestial mechanics, fluid mechanics, and  charged particle dynamics. Important resonant effects that control transport in the phase space in such systems are resonant scatterings and trappings. For systems with weak diffusive scatterings the transport properties can be described with the Chirikov standard map, and the map parameters control the transition between stochastic and regular dynamics. In this paper we put forward the map for resonant systems with strong scatterings that result in non-diffusive drift in the phase space, and trappings that produce fast jumps in the phase space. We demonstrate that this map describes the transition between stochastic and regular dynamics and find the critical parameter values for this transition.

\section{Introduction}
Resonant phenomena determine dynamics of many systems  with two time scales: {\it fast} periodic or quasi-periodic  motion  and {\it slow} change of parameters of this motion. The classical examples of such systems are asteroid and planetary dynamics (e.g., \cite{Morbidelli02, bookAKN06}), hydrodynamical vortices (e.g., \cite{RomKedar90}), and various plasma systems with wave-particle interaction, e.g., laboratory experiments on inertial confinement fusion (e.g., \cite{DewaldPRL16}), plasma acceleration (e.g., \cite{BenistiPoP12}), space-plasma shocks (e.g., \cite{Wilson07, Krasnoselskikh13}), and planetary radiation belts (e.g., \cite{Thorne13:nature, Kasahara18:nature}).  In this paper we consider systems with multiple passages through a nonlinear resonance, where main resonant phenomena are trappings into resonance and scatterings on resonance.

Dynamics in such systems can be modeled by the  Hamiltonian
\begin{equation}
H = \frac{1}{2}g(\la) \varepsilon^{-1}\left(p-p_R(\la)\right)^2+  {\rm B}(\la){\cal F}(\phi) \label{equation_0}
\end{equation}
Here $(p, \phi)$ are conjugate momentum and fast rotating phase (with the frequency  $\dot \phi\sim 1/\varepsilon$, $\varepsilon\ll 1$),  $\la$ is  a slowly changing parameter (with the rate $\dot \la = O(1)$),  $p_R$,  $g$, and ${\rm B}$ are system characteristics  depending on $\la$, and $|d{\cal F}/d\phi|\leq1$ describes the perturbation of $p, \phi$-dynamics (as a model one can take ${\cal F}=\sin\phi$). The resonance in system (\ref{equation_0}) occurs when the frequency $\dot\phi$ drops to zero (i.e. $p\approx  p_R(\la)$). To describe the resonant dynamics we introduce the new momentum $\P=p-p_R(\la) $ and get  the pendulum Hamiltonian (see Sect. 6.1.7 in \cite{bookAKN06} or \cite{Chirikov59}),
 \begin{equation}
{\cal H} = \frac{1}{2}g(\la) \varepsilon^{-1}\P^2+{\rm A}(\la)\phi+ {\rm B}(\la){\cal F}(\phi) \label{equation}
\end{equation}
where ${\rm A}= (dp_R/d\la)\dot \la$.

For  this Hamiltonian around the resonance, $\P=O( \varepsilon^{1/2})$, variables    ($\P/\eps^{1/2}, \phi$)  change  fast, at the rate $\sim \eps^{-1/2}$,  and parameter   $\la$  changes  slowly, at the rate $\sim 1$.  Depending on the shape of phase trajectories  plotted for frozen $\la$, the resonant systems can be formally separated into two classes: with all resonant trajectories passing through the resonance $\P=0$ (${\rm |B|}<{\rm |A|}$, see Fig. \ref{fig1}(a)) and with some resonant trajectories oscillating around the resonance (${\rm |B|}>{\rm |A|}$, see Fig. \ref{fig1}(b)). The area $S\sim \sqrt{\varepsilon}$ of the region filled with such closed trajectories is the key characteristics of the resonant systems (see Sect. 6.4.7 in \cite{bookAKN06}).

We consider the situation when parameter $\la$ changes periodically, and the system repeatedly passes through the resonance. During the period of change of $\la$ the system passes through resonance several times. However, for simplicity of the exposition, we will assume that only one such a passage has a considerable effect on dynamics  (value ${\rm B}$ is small at the other passages). Thus, at the resonance we can  use $p=p_R(\la)$ relation and consider coefficients in (\ref {equation}) as functions of $p$ (e.g., $S=S(p)$). The phase $\phi$ variation on this period is given by the Hamiltonian equation (\ref{equation_0}): $\delta\phi=\varepsilon^{-1}\int{g(\lambda)(p-p_R(\lambda))d\lambda/\dot\lambda}=\tau(p)/\varepsilon$ where $\tau(p)=2\pi(1+k_0p)$ and we use such normalizations of $g$, $p_R$ functions that $\int{g(\lambda)p_R(\lambda)d\lambda/\dot\lambda}=-2\pi$, $\int{g(\lambda)d\lambda/\dot\lambda}=2\pi k_0$ (all integrals include one period of $\la$ change)

Far from the resonance, $p$ stays constant up to small oscillations. Effectively, $p$ changes only due to dynamics in a narrow resonant region where $\P=p-p_R= O( \varepsilon^{1/2})$.  Between two consecutive passages through this region the phase $\phi$ changes by a value $\tau(p)/\eps$.  Let $\phi_R$ be a value of the phase $\phi$ at the first crossing of $\P=0$ during one period of $\la$ change. Passage through the resonance can be  characterized by some new phase $x$ which differs from $\phi_R$ by a $2\pi$-periodic function of $\phi_R$.  This new phase is a normalized value of the Hamiltonian on the phase portraits in  Fig. \ref{fig1} (see details in  \cite{Neishtadt99}).
 In the case of passage through resonance without trapping,  $p$ experiences a change $\delta p(x,p)$ during each passage.  Thus we have a map which in the principal approximation has  a form  $p \mapsto p+\delta p(x,p), x\mapsto x+\tau(p)/\varepsilon$ \cite{Lichtenberg&Lieberman83:book, bookSagdeev88}. Here phase $x$ is defined ${\rm mod}\, 2\pi$,  $ \delta p(x,p)$ is $2\pi$-periodic in $x$, and $\delta p(x,p) \sim \eps^{1/2}$ \cite{Neishtadt99}.

For  systems of the first class (Fig. \ref{fig1}a) there is no average drift $\langle \delta p \rangle_x=0$ \cite{Neishtadt99}. Dynamics of $(x,p)$ in such systems is described by diffusive-like scattering (with the diffusion rate $\sim \langle (\delta p)^2 \rangle_{x}$; $\langle ... \rangle_x$ is the averaging over phase values $x$ for an ensemble of trajectories with the same initial $p$), whereas the map in $(x, p)$ can be  reduced to a map that is similar to the standard Chirikov map \cite{Chirikov79}.

Systems of the second class (Fig. \ref{fig1}b) are characterized by competition of drifts $\langle \delta p \rangle_x=\Delta p_{scat}=-S(p) /2\pi \sim \sqrt{\varepsilon}$ and rapid jumps induced by trappings \cite{Neishtadt99, Dolgopyat12}. Note $\Delta p_{scat}$ is called {\it scattering} despite it is a deterministic shift, because for the equation (\ref{equation_0}) the scattering  $\delta p(x,p)$  is a sum of its mean value $\Delta p_{scat}= \langle \delta p \rangle_x\sim \eps^{1/2}$ and a term of the same order with the average equal to zero. When many passages through the resonance are considered, the effect of the latter term is small and is neglected here.

A trapping (with the following release from the resonance) results in a significant change of $p$,  $\Delta p_{trap} \sim 1$, but only a small fraction of phase points from $(x, p)$ plane with $x\in[0,x^*(p)]$,  $x^*(p)= dS/dp\sim \sqrt{\varepsilon}$  undergo trappings \cite{Neishtadt99}. The change $\Delta p_{trap}$ of $p$ between a trapping and the release is determined by conservation of the adiabatic invariant $\oint \P d\phi$ for trapped particles, i.e. a trapping and the following release occur at the same value of area $S$, see Fig. \ref{fig1}(c).  This fine balance between trappings (and jumps) of a small fraction $\sim \sqrt{\varepsilon}$ of phase points and drifts (with the rate $\sim \sqrt{\varepsilon}$) of untrapped points defines  dynamics on long time intervals including many passages through resonances  \cite{Artemyev16:pop:letter}.

The mapping technique (i.e. construction of the map in $(x,p)$ plane) is well developed for resonant systems with $\langle \delta p \rangle_x=0$, where this technique provides an effective tool for modeling the long-term system dynamics and calculating the critical system parameters that control the transition between stochastic and regular dynamics \cite{Lichtenberg&Lieberman83:book, bookSagdeev88}. However, there is no map for the resonant systems shown in Fig. \ref{fig1}b, i.e. for systems with $S\ne 0$. In this paper we propose a map that describes trapping and scattering, and study the transition between regular and stochastic dynamics in such systems.

\begin{figure}
\centering
\includegraphics[width=18pc]{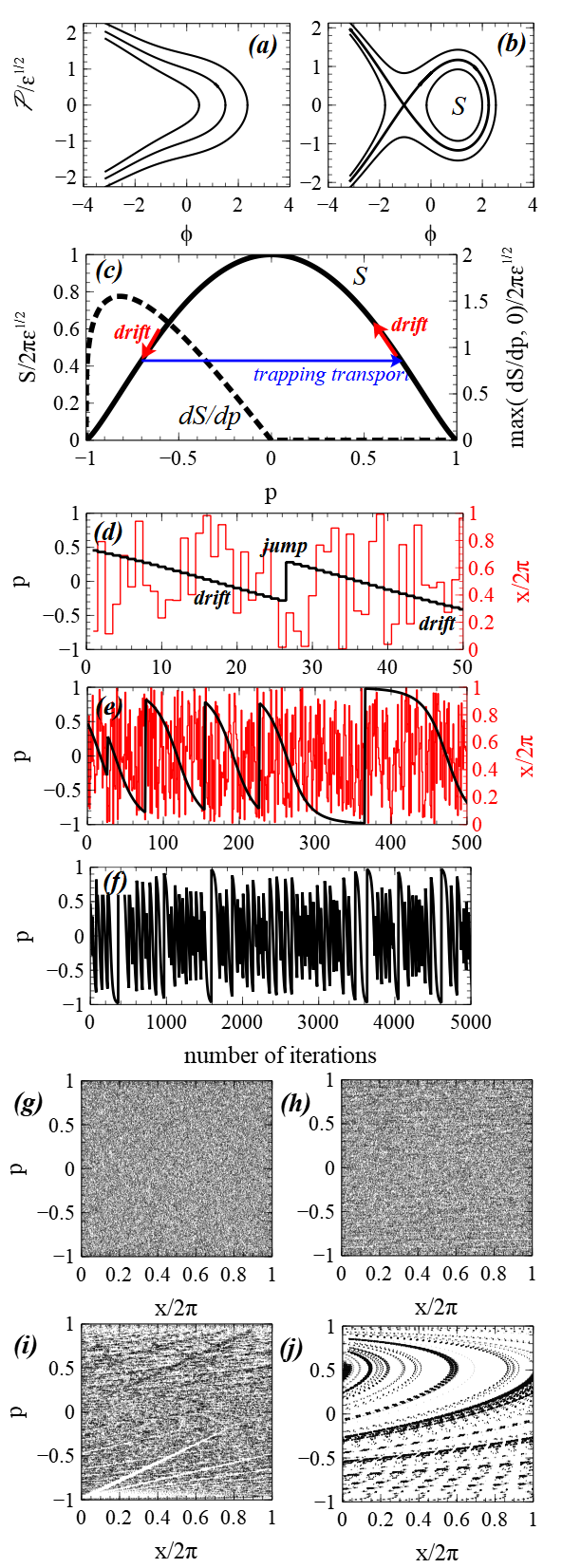}
\end{figure}
\begin{figure}
\caption{Panels (a) and (b) show phase portraits of systems with  diffusive resonant scatterings and with trappings. Panel (c) shows the model $S(p)$ and $\max(dS/dp, 0)$ profiles and explains $p$-dynamics. Panels (d, e, f) show $p$-dynamics on three time scales in the system described by the map (\ref{map}) with $\varepsilon\sim 10^{-3}$. Panels (g-j) show results of $10^6$ iterations of map (\ref{map}) with $\varepsilon\sim 10^{-3}$ and $k_0\sim 10^{-1}$ (g), $k_0\sim 10^{-2}$ (h) , $k_0\sim 10^{-3}$ (i) , $k_0\sim 10^{-4}$ (j).}
\label{fig1}
\end{figure}

\section{System dynamics}

Figure \ref{fig1}(c) shows an example of $S(p)$ profile and directions of trapping jumps and scattering drifts. For each $p<0$ there is a certain range of $x\in[0, x^*]$ from which the phase point will be trapped, whereas for $x\in(x^*,2\pi]$ the phase point will be scattered with $p$ change $\Delta p_{scat}=-S/2\pi$. We choose model $S(p)=2\pi\sqrt{\varepsilon}(1-p^2)^{5/4}$ that satisfies the asymptotic  behavior of small $S$ around $p=\pm 1$ (see Appendix in \cite{Artemyev18:physd}).
  For $p<0$, the map $p \mapsto \bar p, x\mapsto \bar x$ describing the system in Fig. \ref{fig1}(c) can be written as
\begin{eqnarray}
 \bar p &=& p + \left\{ {\begin{array}{*{20}c}
    { - S(p)/2\pi,} & {x \in \left( {\left. {x^* (p),2\pi } \right]} \right.}    \nonumber\\
   { - 2p,} & {x \in \left[ {0,x^* (p)} \right]}  \nonumber \\
\end{array}} \right. \nonumber \\
 \bar x &=& x + \frac{{\tau (\bar p)}}{{\varepsilon }},\quad x^*  =  \frac{{dS}}{{dp}}
\label{map}
\end{eqnarray}
where $\tau(p)=2\pi(1+k_0p)$  and $k_0$ controls the inhomogeneity  of  $\tau$ (see details below). For  $p>0$ the map in $p$ is  $\bar p=p - S(p)/2\pi$ for any $x$.  The ratio  $k_0/\varepsilon$ controls the system dynamics.

The map (\ref{map}) is based on two important relations between scattering and trapping effects: (1) the probability of a trapping $x^*/2\pi$ (i.e., the relative measure of $x$ values for which trappings occur) equals to the gradient of the $p$ scattering rate $-S(p)/2\pi$; (2)  a trapping and the consequent release  occur at the same  value of $S$ (transition $\bar p = -p$ is due to the symmetric shape of $S(p)$ chosen for simplicity). These two relations guarantee the conservation of the phase space volume (absence of dissipation) in systems with trappings and scatterings (see details in \cite{Artemyev16:pop:letter, Artemyev18:physd}).

Map (\ref{map}) describes changes of $p$  composed of the drift to smaller $p$ (with steps $-S/2\pi \sim\sqrt{\varepsilon}$) and jumps to larger $p$ (from $-|p|$ to $|p|$) with a small probability of such jumps. Figure \ref{fig1}(d) shows a fragment of $p$ and $x$ evolution including one such jump. The long term $p$-dynamics includes many jumps connected by drifts (see Fig. \ref{fig1}(e,f)). For large $k_0/\varepsilon$ this long-term dynamics uniformly fills the entire $(x, p)$ plane (see Fig. \ref{fig1} (g)). However, Fig. \ref{fig1} (h, i) shows that with decrease  of $k_0/\varepsilon$ regions of more/less dense concentration of phase points appear (i.e., phase space becomes structured). For $k_0/\varepsilon \leq 1$ the phase space is filled with regular structures (see Fig. \ref{fig1}(j)) indicating  a regular phase point motion (see for comparison \cite{Lichtenberg&Lieberman83:book, bookSagdeev88}). Therefore, there is a critical $k_0^*(\varepsilon)$ value separating regular and stochastic motions for the map (\ref{map}).

\section{Correlation decay}

\begin{figure}
\centering
\includegraphics[width=18pc]{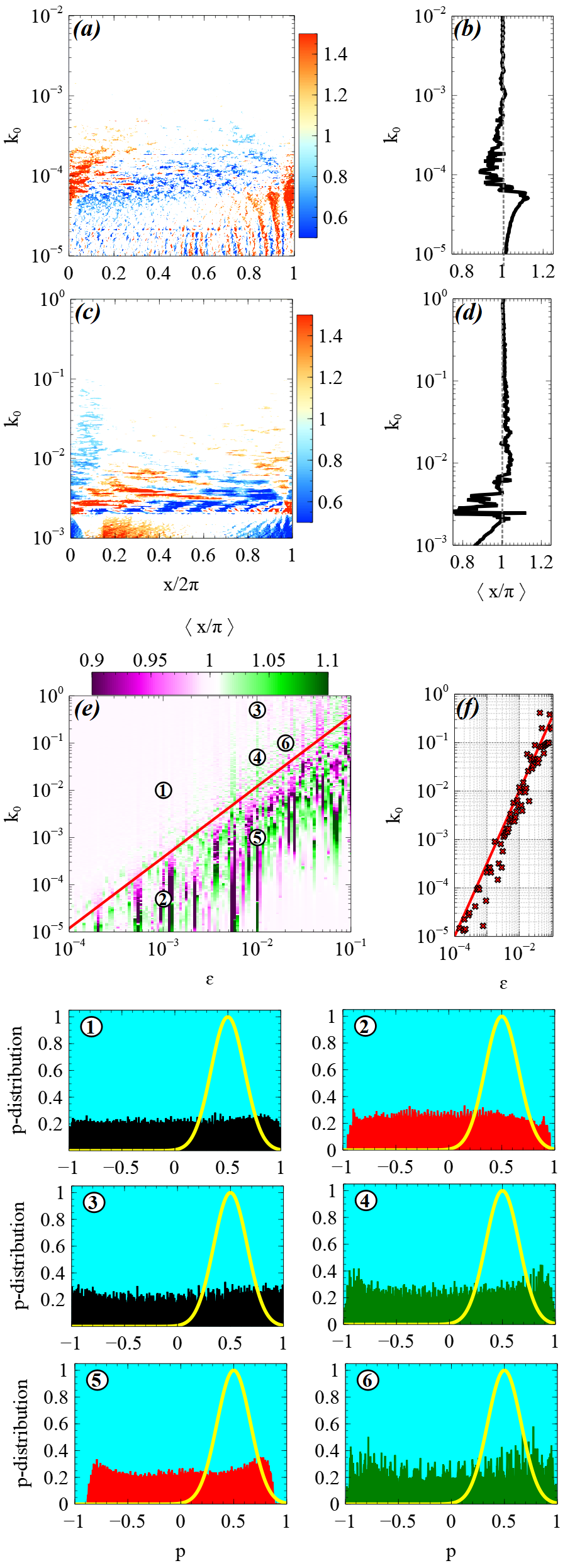}
\end{figure}
\begin{figure}
\caption{Panels (a) and (c) show $x$-distributions for $100$ values of $k_0$ (distributions are normalized in such a way that the uniform distribution equals to one for all $x$), and panels (b) and (d) show averaged $\langle x/\pi \rangle$ as a function of $k_0$ ($\varepsilon\sim 10^{-3}$ for (a,b) and $\varepsilon\sim 10^{-2}$ for (c,d)). Panel (e) shows $\langle x/\pi \rangle-1$ in $(k_0, \varepsilon)$. Panel (f) shows $k_0^*(\eps)$  dependence calculated from panel (e): for $k_0>k_0^*$ the deviation $|\langle x/\pi \rangle-1|<0.05$. The red line is $k_0^*=10\cdot\varepsilon^{3/2}$. Numbers in panel (e) show $(k_0, \varepsilon)$ parameters used for $p$-distribution from bottom panels (the initial $p$-distribution is in yellow).}
\label{fig3}
\end{figure}

To estimate the critical $k_0^*$, we consider a set of systems with different $k_0$ and fixed $\varepsilon$. For each system we perform $10^6$ iterations and plot the distribution of $x$. Such distributions obtained for different $k_0$ can be combined into 2D distribution in the $(x, k_0)$ plane. Figure \ref{fig3}(a) shows this 2D distribution for $\varepsilon \sim 10^{-3}$ (for each $k_0$ the $x$-distribution is transformed to a distribution on the interval $(0,1)$, and thus the uniform distribution has the density equals one for all $x$). For large $k_0$, $x$-distributions are uniform (white color in the figure), but with $k_0$ decrease the $x$-distribution becomes more structured (red and blue colors  show deviations from the density of the uniform distribution, i.e. from one). This transition between the uniform $x$-distribution (i.e., stochastic $x$-dynamics) to more structured one (i.e., more regular $x$-dynamics, but still including elements of chaotic behavior) occurs when the correlation of $x$ and $\bar x$ becomes essential. This change of $x$-distributions with decrease of $k_0$  is well seen in Fig. \ref{fig3}(b), where we plot the average $\langle x/\pi \rangle$ as a function of $k_0$ ($\langle... \rangle$ is the averaging over $x$-distribution with $k_0$ fixed). The value of $\langle x/\pi \rangle$ is close to one for uniform $x$-distributions, but deviates from one as soon as structures appear in the $x$-distributions. Therefore, we can use $\langle x/\pi \rangle-1$ as a measure of deviation of the $x$-distribution from the uniform one. Figures \ref{fig3}(c,d) show the same distribution for $\varepsilon \sim 10^{-2}$:  with larger $\varepsilon$ the $x$-distribution deviates from the uniform one at larger $k_0$.

Figure \ref{fig3}(e) shows $\langle x/\pi \rangle-1$ as a function of $k_0$ and $\varepsilon$. There is the clearly seen area of $\langle x/\pi \rangle\approx 1$ (white color) that corresponds to the uniform $x$-distributions (absence of $x$ and $\bar x$ correlations). This area occupies ranges of large $k_0$ and small $\varepsilon$. The increase of $\varepsilon$ (or decrease of $k_0$) leads to deviation of $\langle x/\pi \rangle$ from one (colored area). To determine the boundary of this area, for each $\varepsilon$ we evaluate $k_0$ corresponding to $|\langle x/\pi \rangle-1|>0.05$ and plot these $k_0(\varepsilon)$ points in Fig. \ref{fig3}(f). The obtained dependence can be fitted by $k_0^*\approx 10\cdot\varepsilon^{3/2}$ (shown in panels (e,f)), i.e., numerical calculations gives the scaling $k_0^* \sim \varepsilon^{3/2}$.

The transition of $x$-dynamics from stochastic to more regular should result in similar transition of $p$-dynamics. To show this effect, we consider six pairs of $k_0, \varepsilon$ parameters. For each pair we calculate $10^4$ trajectories with different initial $x, p$ and $10^6$ iterations. Then we plot $p$-distributions together with the initial $p$-distribution given by $\exp(-16(2p-1)^2)$. Bottom panels of Fig. \ref{fig3} show six $p$-distributions and corresponding locations of the system in the $(k_0, \varepsilon)$ plane. There are two $p$-distributions for $k_0>k_0^*$ (shown by black color), two $p$-distributions for $k_0\sim k_0^*$ (shown by green color), and two $p$-distributions  for $k_0< k_0^*$ (shown by red color). The stochastic $x$-dynamics for $k_0>k_0^*$ results in stochastic $p$-dynamics, and final $p$-distributions are almost uniform, i.e. phase points are uniformly distributed in the $(x, p)$ plane. For $k_0\sim k_0^*$ the $p$-distributions deviate from the uniform one, but still fill the entire $p$ range. These $p$-distributions (shown by green) contain some structures that cannot be seen in the black $p$-distributions. More regular (or less stochastic) $x$-dynamics at $k_0<k_0^*$ results in more regular $p$-dynamics, and the corresponding $p$-distributions (shown in red) do not cover the entire $p$ space. Therefore, $k_0^*\sim \varepsilon^{3/2}$ defines the parametric boundary that separates stochastic and regular dynamics in the $(x, p)$ plane.

\section{Discussion and conclusions}

Although the map (\ref{map}) describing the dynamics in systems with drifts and trapping is quite different from the classical maps describing the diffusive-like scattering \cite{Chirikov79, Lichtenberg&Lieberman83:book, bookSagdeev88}, this map has the same property of transition between stochastic and regular dynamics. This transition occurs when the rates of nominal slow and fast motions become comparable, i.e. when $d\tau/dp \sim \varepsilon^{3/2}$. For systems with $k_0>k_0^*(\varepsilon)$ the consecutive scatterings can be considered as independent $(x, p)$ changes, and thus for such systems the Fokker-Planck-type equations for $p$-distribution can be derived \cite{book:VanKampen03}. For systems with $k_0<k_0^*(\varepsilon)$ the consecutive scatterings are correlated, and such correlations prevent the application of classical kinetic equations for description of $p$-distribution evolution. This conclusion limits the use of the random phase approximation to describe the long-term dynamics in various systems with nonlinear resonant interaction (trapping and scattering drift). Thus construction of Fokker-Planck-type equations for such systems (e.g., \cite{Omura15, Artemyev18:physd}) should be constrained to the parameter ranges with $k_0>k_0^*(\varepsilon)$.

To conclude, we have proposed the map describing dynamics in a resonant system with the effects of drifts and trappings. This map can be used to model long-term dynamics in such systems instead of the approach based on solving a kinetic (Fokker-Planck-type) equation for $p$-distributions (e.g., \cite{Omura15, Artemyev16:pop:letter}). The main advantages of the map (\ref{map}) approach are simplicity of inclusion of additional resonances (i.e., construction of $\Delta p_{scat}$, $\Delta p_{trap}$ as a combination of several terms) and description of the phase-correlation effects.


\bibliographystyle{plain}%

\end{document}